# Delayed Choice, Complementarity, Entanglement and Measurement

Hervé Zwirn[1][2][3]
herve.zwirn@gmail.com

## ABSTRACT

It is well known that Wheeler proposed several delayed choice experiments in order to show the impossibility to speak of the way a quantum system behaves before being detected. In a double-slit experiment, when do photons decide to travel by one way or by two ways? Delayed choice experiments seem to indicate that, strangely, it is possible to change the "decision" of the photons until the very last moment before they are detected. This led Wheeler to his famous sentence: "No elementary quantum phenomenon is a phenomenon until it is a registered phenomenon, brought to a close by an irreversible act of amplification". Nevertheless some authors wrote that backward in time effects were needed to explain these results. I will show that in delayed choice experiments involving only one particle, a simple explanation is possible without invoking any backward in time effect. Delayed choice experiments involving entangled particles such as the so called quantum eraser can also be explained without invoking any backward in time effect but I will argue that these experiments cannot be accounted for so simply because they rise the whole problem of knowing what a measurement and a collapse are. A previously presented interpretation, Convivial Solipsism, is a natural framework for giving a simple explanation of these delayed choice experiments with entangled particles. In this paper, I show how Convivial Solipsism helps clarifying the puzzling questions raised by the collapse of the wave function of entangled systems.

## RESUME

Il est bien connu que Wheeler a suggéré plusieurs expériences dites "à choix retardé" pour montrer l'impossibilité de parler de la manière dont un système quantique se comporte avant qu'il ne soit détecté. Dans une expérience à deux fentes, quand les photons décident-ils de voyager par un seul chemin ou par les deux ? De manière étrange, dans une expérience à choix retardé il semble qu'il soit possible de changer la "décision" du photon jusqu'au dernier moment avant qu'il ne soit détecté. Ceci amena Wheeler à dire : "Aucun phénomène quantique élémentaire n'est un phénomène tant qu'il n'a pas été enregistré par l'intermédiaire d'un acte irréversible d'amplification". Cependant certains auteurs ont écrit que des effets rétroactifs dans le temps étaient nécessaires pour expliquer ces résultats. Je montre que dans les expériences à choix retardés n'impliquant qu'une seule particule, il est possible de donner une explication simple ne nécessitant aucun effet rétroactif dans le temps. Les expériences impliquant des particules intriquées, comme "la gomme quantique", peuvent également être expliquées sans aucun recours à des effets rétroactifs dans le temps mais je défendrai l'idée selon laquelle on ne peut pas en rendre compte aussi simplement car elles soulèvent le problème de savoir ce que sont exactement une mesure ou une réduction de la fonction d'onde. Une interprétation que j'ai déjà proposée, le Solipsisme Convivial, fournit un cadre naturel pour donner une explication claire de ces expériences à choix retardés avec des particules intriquées. Dans cet article, je montre comment le Solipsisme Convivial permet de clarifier les questions posées par la réduction de la fonction d'onde de systèmes intriqués.

---

[1] LIED, UFR de Physique, Université Paris 7, CNRS, Paris France
[2] CMLA, ENS-Cachan, CNRS, Université Paris-Saclay, Cachan, France
[3] IHPST, CNRS, Université Paris 1, ENS, Paris, France

## 1. Introduction

In his notorious discussion of the double-slit experiment [1], Feynman claims that this very experiment exemplifies the main mystery of quantum mechanics. It is intended to make explicit the complementarity between a "wave-like" behavior attributed to photons when an interference pattern is observed and a "particle-like" behavior when there is no interference. It is true that complementarity is one of the most important differences between quantum mechanics and classical mechanics. Properties that are considered as simultaneously observable features of classical systems, such as for example position and momentum, are no longer simultaneously measurable for quantum systems. Worse, this complementarity is not limited to position and momentum but is the general rule for all non-commuting observables. Many Gedankenexperiments have been designed to illustrate these features. A lot of them have been really done and the predictions of quantum mechanics have always been confirmed.

In the standard presentation that is most often given of the double-slit experiment, either the slit the photon passes through is not observed and one can see an interference pattern on the screen (this is interpreted as the fact that the photon went simultaneously through both slits) or detectors are placed in front of the slits, and there is no interference because the photon is supposed to have gone through only one slit (the one indicated by the detector which clicks). These two situations correspond to two mutually exclusive experimental devices because it is not possible to detect the photon without destroying it. That is a perfect illustration of complementarity and the initial explanation given by Bohr many years ago is that the disappearance of the interference pattern, when the slit the photon passes through is observed, must be attributed to the disturbance caused to the photon by this very observation, in accordance with the uncertainty principle.

This led Wheeler [2] to propose a set of so called "delayed choice" experiments to try to analyze when exactly the photon "decides" which behavior it is going to adopt. The choice of the experimental device that is used is decided at the very last moment before the measurement. In the above presentation the photon seems to be able to choose which behavior it is going to adopt depending on the whole experimental setup that is in place, which is already strange enough. But in the case of a delayed choice, it is even stranger since it seems that changing the experimental device at the very last moment has an effect on what the photon decided well before, requiring a sort of backward in time communication!

Many objections can be done against this presentation and against Bohr's explanation. First, as we will see, the behavior that we have supposed the photon adopts immediately after the slits (gone through both slits in case of interference or only through one if one detector clicks) is a misleading interpretation.

Second, new experiments in which it is possible to know which slit the photon passes through without in any way disturbing the photon [3, 4, 5, 6, 7, 8, 9] show that Bohr's explanation is not satisfying[4].

But this last kind of experiments involves more than one particle. That shows that Feynman was wrong thinking that the only mystery of quantum mechanics lies in complementarity. One probably even more puzzling feature of quantum mechanics is entanglement. Experiments where entangled particles are concerned raise more difficult questions than experiments where only one particle exhibits a complementary behavior.

In order to clarify these points I will successively analyze different experiments by order of complexity depending on the fact that they involve only one single particle or two entangled ones and that delayed choice is considered or not. First, I will analyze the double-slit experiment and the Mach-Zender interferometer without delayed choice, then with delayed choice. Second, I will switch to experiments with entangled particles such as the delayed choice quantum eraser.

For sure, a great many explanations given to these experiments involve the question of measurement and of wave packet reduction or collapse. The apparently instantaneous collapse is often presented as creating a tension with Relativity and can even in the EPR situation seem to raise problems with causality. In case of delayed choice, the tension is more drastic since some authors see there a possibility to have a backward in time causality [12] or to require advanced waves [13]. So a proper analysis of these experiments demands to be very clear on what a measurement is and when a collapse occurs (provided a collapse occurs …). I will expose how, using the Convivial Solipsism interpretation [14, 15, 16] and the solution to the measurement problem that I proposed in a recent paper [16], it is possible to avoid these difficulties and to understand the EPR paradox and the delayed choice quantum eraser experiments which can be seen as a generalization of the EPR situation.

## 2. Delayed choice experiments without entanglement
### 2.1. Usual presentation

Wheeler [2] designed a set of delayed choice experiments that were only thought experiments when he proposed them but which have been done for some of them at the present time. I will start by giving a description of these experiments similar to the way they are often presented by their authors or by some commentators. The essence of all these experiments can be presented through a double slit experiment where the choice between either detecting the slit the photon went through or not detecting it can be done at the very last moment (see figure 1).

---

[4] We will not discuss here the experiments [10, 11] attempting to give a meaning to the trajectories of particles as is the case inside the Bohmian interpretation of quantum mechanics.

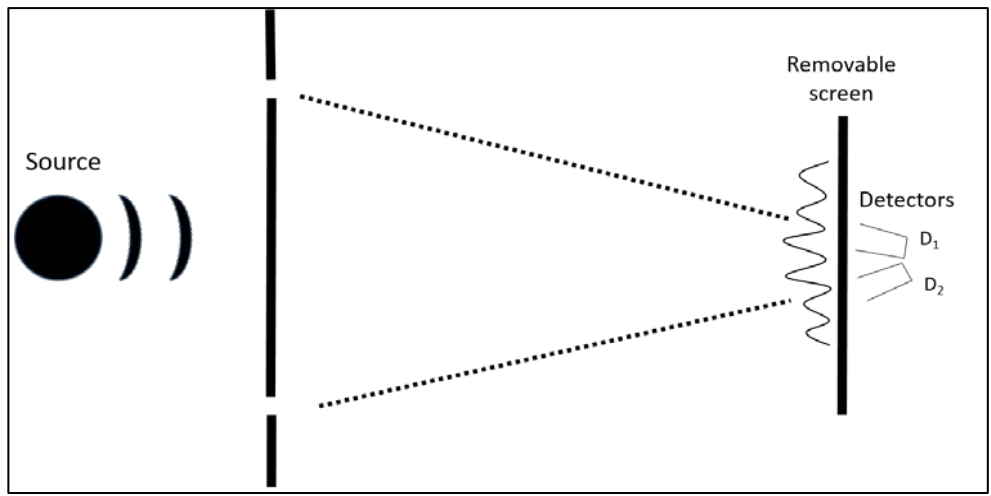

**Fig. 1: Delayed choice double slit experiment**

A source sends photons one by one. Either a screen (removable) receives the light or two detectors ($D_1$, $D_2$), each one pointing precisely at one slit, receive it. If the screen is present, then after a while, an interference pattern is seen. If the screen is removed, for each photon only one detector clicks allowing to know which slit the photon went through. The usual presentation says that when the screen is present (interference), each photon decides to pass through both slits while when it is removed (no interference), each photon chooses only one slit to pass through. A strange but possible interpretation would be that the photon is "informed" of the presence or of the absence of the screen (no matter how far away it is) when it passes the slits. A delayed choice situation corresponds then to the case where choosing to put or remove the screen is done well after the photon went through the slits. In this case, the conclusion seems to be that putting or removing the screen has an influence in the past on the decision of the photon to go through only one or both slits.

In a similar experiment, a single photon is split by a first beam splitter $BS_1$ of a Mach-Zender interferometer into two beams which are reflected by mirrors A and B to a crossing point C (see figure 2). If a second beam splitter $BS_2$ is present at C (closed configuration), then the two beams recombine producing interference and if two detectors $D_1$ and $D_2$ are present, only $D_1$ will click (destructive interference at $D_2$). If there is no second beam splitter at C (open configuration), then $D_1$ or $D_2$ can click with equal probability. In the second case, we know by which route the photon came and this prevents any interference to happen while in the first case, the interference shows that the photon came by both routes. As it is impossible to do both experiments at once, this is an illustration of the complementarity between the "wave-like" and "particle-like" behavior of photons. Each one of these incompatible experimental devices allows to highlight one of these two aspects. If the experimental device "closed" is used, the photon will exhibit its "wave-like" attribute which is usually interpreted as if the photon had decided to travel by both routes, while if the experimental device "open" is used, the photon will exhibit its "particle-like" attribute which is usually interpreted as if the photon had decided to choose only one route to travel.

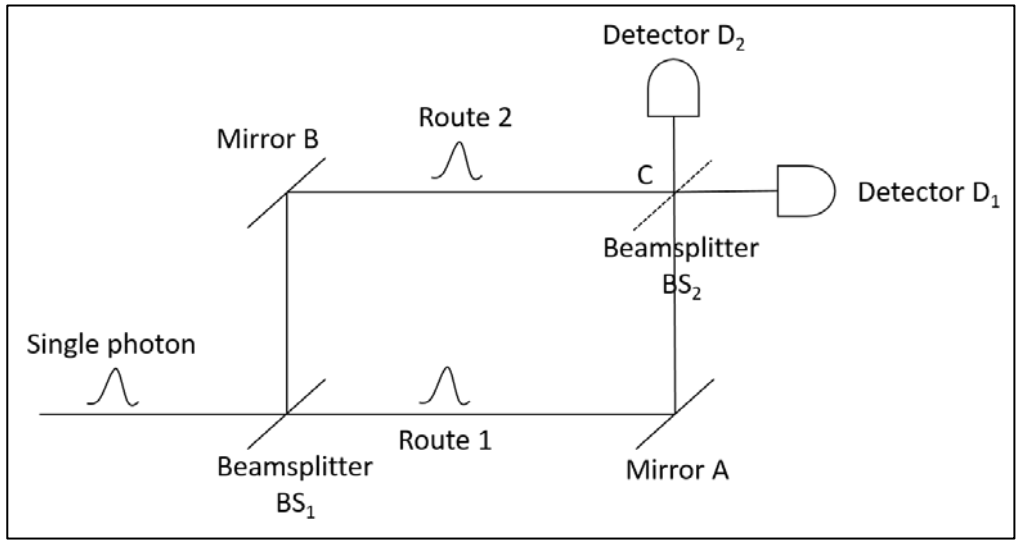

**Fig. 2: Delayed choice Mach-Zender interferometer**

Under this point of view, it would nevertheless be possible to think that the photon, "knowing" or "informed of" which experimental device is used, decides to behave as a particle in case of the "open" configuration or as a wave in case of the "closed" one. Hence, the photon would decide if it is going to travel by one route or by both routes when it meets the first beam splitter. Of course, this is already strange enough as we saw above for the double slit experiment. But Wheeler's delayed choice experiment seems to make things even stranger since one can decide whether to put the second beam splitter or to take it out at the very last moment before the photon arrives at C. So this decision comes after the moment when the photon is supposed to have chosen if it is going to travel by one route or both. This decision comes when the photon has already done a large part of its travel. How then is it possible that putting the second beam splitter or removing it can change what the photon has already done in the past?

This last experiment has been done recently [17] following exactly Wheeler's original idea demanding to use a single particle quantum state and a space-like separation between the choice of the performed measurement and the entering of the particle into the interferometer. And the above result is well confirmed. This leads the authors to say:

*"Our realization of Wheeler's delayed choice GedankenExperiment demonstrates beyond any doubt that the behavior of the photon in the interferometer depends on the choice of the observable which is measured, even when that choice is made at a position and at a time such that it is separated from the entrance of the photon in the interferometer by a space-like interval. In Wheeler's words, since no signal*

*travelling at a velocity less that of light can connect these two events, "we have a strange inversion of the normal order of time". […] Once more, we find that Nature behaves in agreement with the predictions of Quantum Mechanics even in surprising situations where a tension with Relativity seems to appear."*

But is it true? As rightly noticed by Roussel and Stefan [18], we are going to show that this presentation is based on a wrong interpretation of what happens.

### 2.2. What is wrong with this point of view?

To see what is wrong in the previous reasoning, let's come back to the double-slit experiment (figure 1). Following the very clear paper of Gaasbeek [19], let's write the wave function of the photon immediately at the right of the slits. It is the sum of two spherical waves:

$$\psi(\vec{r}) = \frac{1}{|\vec{r}-\vec{r_1}|}e^{ik|\vec{r}-\vec{r_1}|} + \frac{1}{|\vec{r}-\vec{r_2}|}e^{ik|\vec{r}-\vec{r_2}|} \qquad (1)$$

Where k is the wavenumber of the photon and $\vec{r_1}$ and $\vec{r_2}$ are the locations of the two slits. Now:

$$|\vec{r}-\vec{r_2}| - |\vec{r}-\vec{r_1}| \sim d\sin\theta \qquad (2)$$

Where d is the distance between the two slits and $\theta$ the angle of the position $\vec{r}$ on the screen. Then:

$$\psi(\vec{r}) \sim \frac{1}{|\vec{r}-\vec{r_1}|}e^{ik|\vec{r}-\vec{r_1}|}(1 + e^{ikd\sin\theta}) \qquad (3)$$

So constructive interferences arise at points on the screen such as $kd\sin\theta = 2n\pi$ and destructive interferences at points such as $kd\sin\theta = (2n+1)\pi$. This confirms the wave-like behavior and the fact that the photon went through both slits. Now, if the screen is removed, remembering that the detectors measure the momentum of the incoming photon, with a very narrow detection range, it is possible to write:

$$\psi(\vec{r}) \sim \frac{1}{|\vec{r}-\vec{r_1}|}e^{i\vec{k_1}(\vec{r}-\vec{r_1})} + \frac{1}{|\vec{r}-\vec{r_2}|}e^{i\vec{k_2}(\vec{r}-\vec{r_2})} \qquad (4)$$

Where $\vec{k_1}$ and $\vec{k_2}$ are wavevectors of size *k* directed along $\vec{r_T}-\vec{r_1}$ and $\vec{r_T}-\vec{r_2}$ respectively and $\vec{r_T}$ is the location of the detectors. When detector D₁ clicks, the momentum is around $\vec{k_1}$ and when detector D₂ clicks, the momentum is around $\vec{k_2}$. Let's imagine the two detectors as belonging to an apparatus A built in such a way that when the spherical wave interacts with it, a correlation is established between the momentum is around $\vec{k_1}$ and the detector D₁ clicks (state $|D_1\ \text{clicks}\rangle$ of the apparatus A) and between the momentum is around $\vec{k_2}$ and the detector D₂ clicks (state $|D_2\ \text{clicks}\rangle$ of the apparatus A). Then we can see D₁ and D₂ as similar to the two possible positions of the needle of an apparatus measuring a two valued momentum observable.

So, when the photon interacts with this apparatus, the global wave function of the photon plus the apparatus can be written:

$$\psi(\vec{r}) \sim \frac{1}{|\vec{r}-\vec{r_1}|} e^{i\vec{k_1}(\vec{r}-\vec{r_2})} |D_1 \text{ clicks}\rangle + \frac{1}{|\vec{r}-\vec{r_2}|} e^{i\vec{k_2}(\vec{r}-\vec{r_2})} |D_2 \text{ clicks}\rangle \qquad (5)$$

Then, according to the Born rule, the respective probabilities that $D_1$ or $D_2$ click are proportional to $\frac{1}{|\vec{r}-\vec{r_1}|}$ and $\frac{1}{|\vec{r}-\vec{r_2}|}$ which are equal. So, $D_1$ and $D_2$ click each one in fifty per cent of cases.

Now, thinking that because the photon has been detected by detector $D_1$, it followed the dotted line from the upper slit to $D_1$ (see figure 1) is false. It is exactly the same mistake than thinking that because the spin along Oz of an electron has been measured to be equal to +1/2, this spin was already +1/2 before the measurement. It is well known that a quantum measurement does not reveal a preexisting state of affairs but "creates" the result (except for the cases where the system is already in an eigenstate of the observable that is measured). Equation (5) shows that before being detected by either of the two detectors, the photon was a spherical wave.

### 2.3. A point of view (almost) without mystery

What we showed above gives a simple way to understand what happens in this kind of experiments, with or without delayed choice (in this case, it does not change anything). The photon has a quantum behavior from the moment it is emitted until the moment it is detected either on the screen or by one detector (i.e. until a measurement is made on it). That means that before its detection, the photon is always in a superposed state described by $\psi(\vec{r})$ in equation (1) (gone through both slits in the case of the double slit experiment or having followed both routes in the case of the interferometer). In the case of the double slit experiment, it is only when it meets the screen (playing the role of an apparatus measuring a continuous position observable) or the two detectors (playing the role of an apparatus measuring a two valued momentum observable) that the wave function of the photon collapses on one of the possible eigenvectors of the observable that is measured. The interference pattern is nothing else than the reflection of the Born rule applied to the measurement of a continuous position of the photon with a superposed wave function $\psi(\vec{r})$. The clicks of the detectors are nothing else than the reflection of the Born rule applied to the measurement of a two valued momentum observable of the photon with the same superposed wave function $\psi(\vec{r})$. Thinking that when an interference pattern is seen on the screen, the photon went through both slits while when a detector clicks, it went through only one, is simply a mistake. It is then clear that putting the screen or the detectors at the very last moment has strictly no importance. A similar reasoning can be made for the experiment with a Mach-Zender interferometer. The photon will always go through both routes after the first beam splitter. Then the wave function of the photon when it arrives on the two detectors will be different if the photon has gone through the second beam splitter or not. If the beam splitter is absent the wave function of the photon at $D_1$ and $D_2$ will give an equal probability for either of the two detectors to click. If the beam splitter is present, the wave function will be changed in passing through it and will give at $D_1$ a probability 1 to click and at $D_2$ a probability 0. But between the first beam splitter and the point C the wave function of the photon (and its behavior) will be strictly the same whether or not the second beam splitter is present.

The conclusion is that Feynman's presentation is misleading because it lets think that on some occasions the photon goes through one slit and on some others it goes through both slits. Actually, it always goes through both and the fact that a detector aimed at one slit clicks must not be interpreted as the fact that the photon "really" went only through this slit before its detection. That is as illegitimate as it would be to interpret a spin + measured along Oz for a spin ½ particle as the fact that the spin was already + before the measurement or to think that a particle follows a continuous trajectory between two successive position measurements. It is even strange that such an interpretation arises in this context since Wheeler himself said [2]:

*"No elementary phenomenon is a phenomenon until it is a registered (observed) phenomenon. It is wrong to speak of the "route" of the photon in the experiment of the beam splitter. It is wrong to attribute a tangibility to the photon in all its travel from the point of entry to its last instant of flight."*

Once this has become clear, the delayed choice adds nothing and of course, no "backward in time" action is needed to explain what happens.

Notice that, in the above presentation, I have used the reduction postulate and the Born rule as if their use was perfectly unproblematic and as if the simple interaction of the photon with a detector was enough to collapse the wave function. Actually, that is not so simple and what I have said ignores deliberately the difficulties raised by the measurement problem itself. But this (remaining) mystery can be left aside for understanding these experiments involving only one particles. We are going to see that, on the contrary, it will be necessary to take it into account if we want to find a proper explanation when entanglement is involved.

3. **Entanglement without delayed choice (EPR situation)**

As is well known, in the EPR experiment [20, 21] one considers a setup where two particles A and B in a singlet state are measured by two spatially separated experimenters Alice and Bob:

$$|\psi\rangle = \tfrac{1}{\sqrt{2}} \left[ |+\rangle^A |-\rangle^B - |-\rangle^A |+\rangle^B \right] \qquad (6)$$

Let's say that Alice does her measurement first. Then Bob will find the opposite outcome even if his measurement is space-like separated from Alice's one. It is usually stated that the collapse caused by the first measurement on A instantaneously results in a collapse of the state vector of B. Thinking that the result is determined from the moment the particles separate is not allowed since Bell's inequality [22] forbids local hidden variables. But if the two measurements are space-like separated no one can be said to be before the other in an absolute way. For two observers moving in the opposite direction, Alice's measurement will be the first for one of them while Bob's measurement will be the first for the other. So, which one of the two measurements causes the result of the other?

To answer this question it is necessary to clarify the position one adopts about the status of the wave function and the collapse, thing which is actually seldom done by those who discuss the problem. An usual way to escape this seeming violation of Relativity is to acknowledge the fact that quantum mechanics is not local but to say that since it is impossible for Alice or Bob to choose the outcome they get, there is no way to use the collapse to communicate instantaneously. That is true but that is not enough to solve the problem if one considers the state vector to represent the real state of the system and the collapse to be a physical event changing the state vector (hence changing the real physical state of the system). If a random number generator that is used at one place was able to cause the apparition of exactly the same random number instantaneously at another location, it would not be possible to use it to communicate but that would nevertheless be considered as violating Relativity. So for many physicists a way to avoid the problem is often to say that in the EPR case, we notice a mere correlation between two results and to claim that correlation is not causality [19]. But this is not an acceptable reason since in classical statistics the precise reason for the difference between correlation and causality is the fact that a common cause can be invoked which is here forbidden by Bell's inequality.

At last, it is easy to prove that in general the order of measurements made on two entangled particles has no impact on the conditional probabilities of these measurements provided that each observable concerns only one particle [19]: $P[(O_A = i)$ and $(O_B = j)]$ is the same whether the measurement of $O_A$ on A is made before or after the measurement of $O_B$ on B. But that does not bring any clarification if the state vector is assumed to represent the real state of the system and if the collapse is supposed to be a real change in this state vector. Abandoning at least a part of realism seems necessary. Some physicists sweep things under the carpet and say that this problem cannot be properly discussed in non-relativistic quantum mechanics. Anyway, no satisfying explanation is given and the question of why, how and when the collapse occurs is usually neglected. We are going to see that this question arises even more acutely in the delayed choice experiment with two entangled particles and I will propose a possible explanation involving a solution of the measurement problem which drastically changes the point of view that we must have on these questions.

## 4. Delayed choice with entanglement
### 4.1. The experimental setup

Let's now switch to some experiments usually known as "delayed choice quantum eraser". In a way, these experiments can be considered as a generalization of the EPR situation. They show that the loss of interference on one particle is due not to the uncertainty principle but to the measurement of the twin entangled particle [3]. Many such experiments have been made [4, 5, 6, 7, 8, 9] but we will here restrict our attention to one of them [4] which contains all is necessary to analyze the subject.

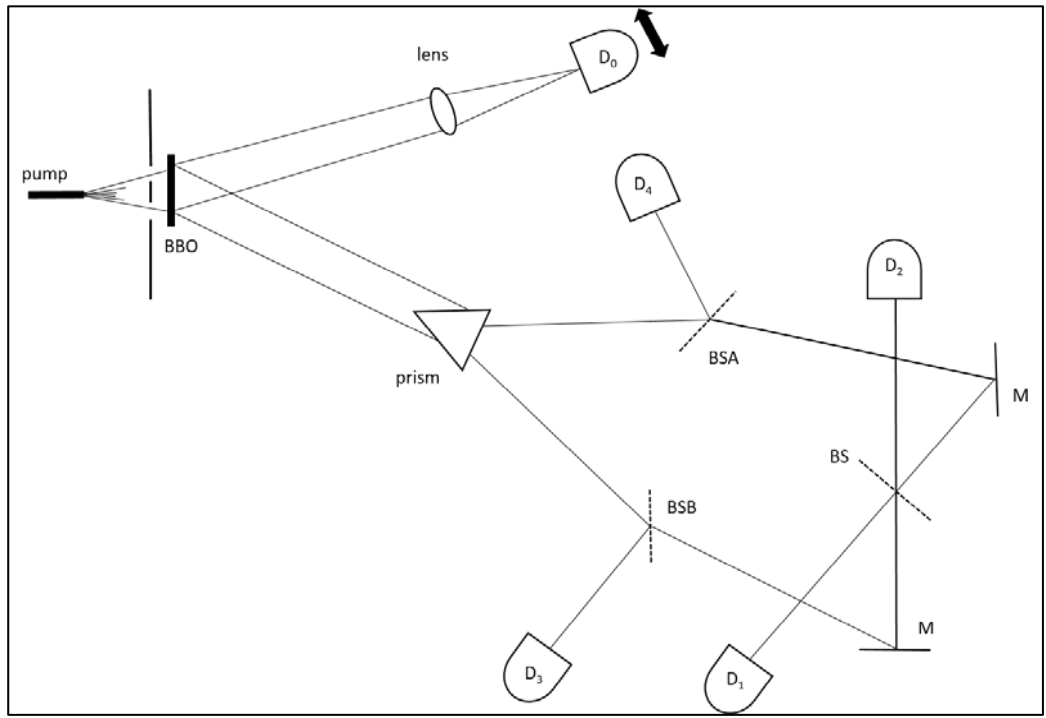

**Fig. 3: Delayed choice quantum eraser**

In this experiment (see figure 3), light is sent through a double slit and goes through a BBO crystal for a parametric down conversion splitting each photon in two entangled photons. $D_0$ is a moveable detector playing a role similar to a screen to detect an interference pattern for the photons travelling upward (signal photons). The photons travelling downward (idler photons) are received by a prism and a set of beam splitters (BS) and mirrors (M). If a photon is detected by $D_1$ or $D_2$, it may have travelled via either of the two routes. If it is detected by $D_3$ or $D_4$, it can only have come through one route. The experimental set up is such that the signal photon is detected by $D_0$ before the detection by one of the other detectors of its idler twin photon.

When the results of the detectors are compared through the joint detection rate (the time delay between a triggering of $D_0$ and a triggering of $D_{i\ (i=1\ to\ 4)}$ is constant), one sees that the signal photons associated with idler photons detected by $D_1$ or $D_2$ show an interference pattern while the others show no interference.

The beam splitter BS is called the eraser because if it is present, it is no more possible to know which slit the photons detected by $D_1$ and $D_2$ came from while if it is taken out, a click of $D_1$ (resp. a click of $D_2$) is associated with a photon having gone through the upper slit (resp. the lower slit) and the interference pattern is loss.

One important point to notice is that it is possible to recognize an interference pattern only through the extraction, from all the detections registered by $D_0$, of those corresponding either to detections by $D_1$ or to detections by $D_2$. That has the consequence that it is impossible to use this device to transmit information from the future to the past because recognizing an interference pattern is possible only when one have the knowledge of which detection by $D_0$ corresponds to which detector $D_i$. In particular, the tempting idea consisting in putting or removing the beam splitter BS in the far future to produce interference (BS present) or not (BS absent) in the present, is not working because there is no way to distinguish an interference pattern from no interference just by looking at $D_0$ since the interference patterns correlated to $D_1$ and to $D_2$ have a $\pi$ phase shift producing jointly exactly the same image than no interference at all. Hence it is impossible to separate them without knowing which impact corresponds to which detector and this necessitates information that can be known only after deciding to put or not the beamsplitter.

### 4.2. Discussion

If we adopt the point of view (that we criticized above) according to which photons showing interference must have gone through both slits and photons showing no interference have gone through only one slit, it seems that the signal photon has to make a choice before its idler twin be detected. So the question is: how is it possible for a signal photon to make a choice depending on what will happen to its idler twin later?

Actually this question rests on many implicit (wrong) assumptions. The first one is that photons showing interference must have gone through both slits while photons showing no interference have gone through only one slit. We have seen before that this is not true. The second one is that in this picture, one reasons as if the detection by $D_{i\ (i=1\ to\ 4)}$ was more important than the detection by $D_0$ because it is assumed that it is the detection by $D_{i\ (i=1\ to\ 4)}$ which decides if there is interference or not (even if this latter detection happens in the future of the detection of interference). But this is a strange asymmetry that is assumed here since nothing tells us that when two particles are entangled, one of them has a superior status such that it is the measurement on it that dictates the result of the measurement on the other even if the measurement on the other is done before. Remember the EPR situation where things are perfectly symmetric and where the order of measurements does not matter since there are cases where it is impossible to say which the first one is. Here also, it could perfectly be possible to design a quantum eraser where the measurements by $D_0$ and by $D_{i\ (i=1\ to\ 4)}$ are space-like separated making impossible to know which is the first in an absolute way.

The reason why this asymmetry is used (wrongly) in this kind of presentation is clear: it allows to develop the argument in order to show that there is an influence of the measurement of the idler photon on the behavior of the signal photon. Knowing which detector $D_1$, $D_2$, $D_3$, or $D_4$ clicked allows to infer in a univocal way that the corresponding signal photon will or will not interfere with itself. On the contrary, acknowledging that there is a perfect symmetry would imply to discuss the case where a detection at $D_0$ is made first. But of course, nothing conclusive on the idler photon can be drawn from such a detection of the signal photon, if only because it is impossible to identify directly signal photons showing interference at $D_0$. Hence, the authors simply forget the symmetry because this is convenient for their argumentation.

The lesson that we got from the EPR situation is that no one of the two measurements can be considered as the "cause" of the result of the other in a physical sense. This is again true in the delayed choice quantum eraser experiment even though the ingenious experimental setup allows to present the argument in a way that hides the symmetry and makes more difficult to escape the intuitive (but misleading) appeal of thinking that it is the fact of knowing or not which slit the idler photon went through that determines the fact that there is interference or not for the signal photon.

One can notice that insofar as it can be shown that the order of the measurements made on two entangled particles has no effect on the values of the conditional probabilities, it is obvious from the beginning (of course if we accept that the predictions of quantum mechanics are true) that all these experiments will provide the same results whether the measurement of the "which path information" on one particle be made before or after the measurement of position on the other. The particles for which the measurement allows to determine the path (more precisely, for which the measurement *seems* to be interpretable as allowing to determine the path) will be associated with twin particles without interference and the particles for which the measurement does not allow to determine the path will be associated with twin particles showing interference, whatever the order of measurements be. But it is wrong to think that it is the fact that the path is determined that causes the loss of interference. This wrong interpretation is the reason why it is sometimes written that there is an influence of the future on the past.

### 4.3. A first attempt of clarification

Following [19], let's write the wave function of the two entangled photons in a simplified form. Right after the slits and before the BBO crystal the wave function of the input photon is a superposition of an upper part and a lower part:

$$|I\rangle \rightarrow \frac{1}{\sqrt{2}} \big[ |U\rangle + |L\rangle \big] \tag{7}$$

Then after the BBO crystal, each component is split into an entangled pair of photons, one going upward and the other going downward:

$$|\psi\rangle \rightarrow \frac{1}{\sqrt{2}}\big[\,|U \nearrow\rangle\,|U \searrow\rangle + |L \nearrow\rangle|L \searrow\rangle\,\big] \tag{8}$$

Then the idler photon meets a beam splitter and arrives to the detectors. Taking the detectors into account in the wave function gives:

$$|\psi\rangle \rightarrow |U \nearrow\rangle\left[\tfrac{1}{2}|4\rangle + \tfrac{1}{2\sqrt{2}}|1\rangle + \tfrac{1}{2\sqrt{2}}|2\rangle\right] + |L \nearrow\rangle\left[\tfrac{1}{2}|3\rangle + \tfrac{1}{2\sqrt{2}}|1\rangle + \tfrac{1}{2\sqrt{2}}|2\rangle\right] \tag{9}$$

Where the coefficients are easy to check on figure 3 and where the states written $|i\rangle$ (for $i = 1$ to 4) stand for the combined states: "photon downward arriving at detector $i$ and detector $i$ clicking".

The state written in equation (9) is exactly analogous (though a bit more complicated) to the singlet EPR state written in equation (6). To make explicit the similitude, let's write it as:

$$|\psi\rangle \rightarrow |U\rangle^{\nearrow}\left[\tfrac{1}{2}|4\rangle + \tfrac{1}{2\sqrt{2}}|1\rangle + \tfrac{1}{2\sqrt{2}}|2\rangle\right]^{\searrow} + |L\rangle^{\nearrow}\left[\tfrac{1}{2}|3\rangle + \tfrac{1}{2\sqrt{2}}|1\rangle + \tfrac{1}{2\sqrt{2}}|2\rangle\right]^{\searrow} \tag{10}$$

We can now compare with equation (6):

$$|\psi\rangle = \frac{1}{\sqrt{2}}\big[\,|+\rangle^A\,|-\rangle^B - |-\rangle^A\,|+\rangle^B\,\big]$$

We see that $\nearrow$ and $\searrow$ stand for the two entangled particles A and B, $|U\rangle$ and $|L\rangle$ stand for $|+\rangle$ and $|-\rangle$ for particle A, while the two more complicated terms into brackets stand respectively for $|-\rangle$ and $|+\rangle$ for particle B. The two equations are similar and all what we said about (6) applies as well to (9) or (10).

First, it is true that the correlations between the measurements on the two particles will be independent of the order in which these measurements are done. So we can reason as if the measurement on the idler photon was done first. To understand what happens in this case, rewrite (10) as:

$$|\psi\rangle = \tfrac{1}{2\sqrt{2}}|1\rangle^{\searrow}\big[\,|U\rangle + |L\rangle\,\big]^{\nearrow} + \tfrac{1}{2\sqrt{2}}|2\rangle^{\searrow}\big[\,|U\rangle + |L\rangle\,\big]^{\nearrow} + \tfrac{1}{2}|3\rangle^{\searrow}|L\rangle^{\nearrow} + \tfrac{1}{2}|4\rangle^{\searrow}|U\rangle^{\nearrow} \tag{11}$$

With the wave function written under this form, it is easy to see that if $D_1$ or $D_2$ clicks, the wave function of the signal photon collapses on $\frac{1}{\sqrt{2}}\big[\,|U\rangle + |L\rangle\,\big]^{\nearrow}$ and this will give rise to interference, while if $D_3$ (resp. $D_4$) clicks, the wave function of the signal photon collapses on $|L\rangle^{\nearrow}$ (resp. $|U\rangle^{\nearrow}$) and there will be no interference.

This "easy" explanation could be considered satisfactory as regards to what happens to each pair of photons depending on which detector clicks. In a way, we have reduced the quantum eraser experiment to a usual EPR experiment.

The problem is that this takes for granted the reduction postulate and the collapse of the wave function. It forgets the questions about what they mean and overlooks the fact that, reassured by the computation showing that the conditional probabilities are the same whatever the order of measurements, we have reversed this order to give an interpretation without scruples while the very physical meaning of this order independence is far from clear.

Notice that all the questions we raised in §3 are still unanswered here. How is it possible that these collapses of entangled particles seem to occur in a time independent way and what is the physical meaning of the correlations they show? What is the physical meaning of the collapse and when does it occur (provided it occurs)? The more spectacular image given by these delayed choice experiments adds actually nothing essential to the main mystery of the non-locality and time independent aspect of the collapse (which are already present in the initial EPR situation which contains, in the sense that Feynman used for the double slit experiment, all the mystery of quantum mechanics) but it emphasizes their oddity. These difficulties lie at the heart of the measurement problem. As I show in a previous paper [16] no satisfying solution has been provided yet (at least inside the standard quantum formalism) and I propose a new solution of the measurement problem through Convivial Solipsism which helps understanding what a measurement is, how the collapse occurs and what is the meaning of these correlations seeming independent of time between entangled particles.

## 5. The measurement problem: Convivial Solipsism
### 5.1. The motivation

In a recent paper [16], I explain why I am not satisfied with the many attempts for solving the measurement problem and I give a list of seven puzzling questions that any interpretation should be able to answer:

1. What is a measurement and when must we use the Schrödinger equation or the reduction postulate for describing the evolution of systems?
2. When and why only one of the many possible results of measuring an observable is selected?
3. If the measurement does not reveal a preexisting value, how is it possible that this value be created during the measurement?
4. Does this value so created belong to the system itself, does it belong to the system and the apparatus, does it concern the external reality and if so, is this reality the same for all observers, or is the value something that concerns merely the observer?
5. If even macroscopic systems can become entangled, why do not we observe macroscopic superpositions?
6. How do we know which observable is measured when we use an apparatus (preferred basis problem)?
7. How must we understand the non-locality shown by Bell's inequalities and is there any instantaneous action at a distance?

I explain how decoherence allows to answer the fifth and the sixth questions and I show that Convivial Solipsism (which welcomes the decoherence mechanism) is an attempt to answer all the others and to get a coherent general image. I have not enough room here to give the details of the reasoning and I refer the reader to [16]. In the present paper, I will briefly state the main features of Convivial Solipsism and I will concentrate on the way it helps understanding what happens in the EPR situation and in the delayed choice quantum eraser experiment which, as seen above, can be considered as a generalization of the EPR situation.

### 5.2. Convivial Solipsism : a brief description

Convivial Solipsism[5] comes from the acknowledgement that it is impossible to give a solution of the measurement problem in a strongly objective way (i.e. without reference to any observer) inside the standard quantum mechanics [23]. Of course this point is still largely debated but I argue in favor of the fact that all the solutions proposed until now are not satisfying [16]. Convivial Solipsism is an attempt to provide a coherent solution at the price of considering that the observer plays a major role in the process of measurement. In Convivial Solipsism, a measurement is not a physical process changing the state of the system. It is the fact for a conscious observer to become aware of a result. This cancels the ambiguity about when to use the Schrödinger equation and when to use the reduction postulate. Convivial Solipsism is based on two main assumptions: the hanging-up mechanism and the relativity of states.

#### 5.2.1. The hanging-up mechanism

Let's consider the standard presentation of the measurement process with an apparatus A, initially in the state $\Psi_A = |A_0\rangle$, of a given observable P on a system S which is in a state that is a superposition of eigenvectors of P, $\Psi_S = \sum c_i |\varphi_i\rangle$. After the interaction between the apparatus and the system, in accordance with the Schrödinger equation, the global system S+A is in an entangled state $\Psi_{SA} = \sum c_i |\varphi_i\rangle |A_i\rangle$ where the state $|A_i\rangle$ of the apparatus is correlated to the eigenstate $|\varphi_i\rangle$ of S associated with the eigenvalue $\lambda_i$ of the observable. The measurement problem comes from the fact that the unitary evolution given by the Schrödinger equation does not allow to get a unique result of the measurement while the reduction postulate says that after the measurement: a) the apparatus is seen in only one of these states, say $|A_{i_0}\rangle$, b) the state of the system is projected onto the related eigenstate $|\varphi_{i_0}\rangle$ and c) the result of the measurement is the eigenvalue $\lambda_{i_0}$ associated with this eigenstate. If the observer is considered, she becomes also entangled with the other systems and according to the Schrödinger equation, the global system after the interaction between the apparatus, the system and the observer is in the entangled state: $\Psi_{SAO} = \sum c_i |\varphi_i\rangle |A_i\rangle |O_i\rangle$ and the same difficulty arises. In the decoherence approach, the environment must also be included but, for sake of simplicity, we omit it here.

---

[5] Convivial Solipsism finds its origin in a criticism of Everett interpretation made by Bernard d'Espagnat [23].

Now, very similarly to what is assumed in Everett interpretation, we make a distinction between the physical brain of the observer and what she perceives. That is a crucial assumption. The physical state of the universe (including the observer and her brain) evolves unitarily through the Schrödinger equation and remains in a superposed state. However the consciousness of each observer can be aware of only one branch of the superposed state and cannot perceive the superposition. The difference with Everett interpretation where there are as many observers perceiving a different result as there are branches is that in Convivial Solipsism there is only one observer whose consciousness hangs-up to only one of the branches through the hanging-up mechanism:

"*Hanging-up mechanism: A measurement is the awareness of a result by a conscious observer whose consciousness selects at random (according to Born rule) one branch of the entangled state vector written in the preferred basis and hangs-up to it. Once the consciousness is hung-up to one branch, it will hang-up only to branches that are daughters of this branch for all the following observations.*"

I refer the reader to [16] for details about the very meaning of the hanging-up mechanism and the way to show that it is consistent with any predictions of various measurements.

### 5.2.2. The relativity of states and the no-conflict theorem

The second assumption is:

"*Relativity of states: Any state vector is relative to a given observer and cannot be considered as absolute.*"

There is no absolute state vector. This is an important point because, together with the hanging-up mechanism, this prevents to keep the traditional image of an external reality which is the same for all the observers. Each observer gets her own state vector for all she is able to observe. This state vector evolves deterministically through the Schrödinger equation and remains always a superposition of states of entangled systems. The physical brain of each observer is part of this universal but relative superposed state and the consciousness of the observer is hung-up to one branch so that the perceptions of the observer are confined to this branch and her daughters when a new measurement is done.

This raises an unavoidable question: can there be any conflict between different observers hung-up to different branches? The answer is no for, as d'Espagnat [23] puts it:

"A*ny transfer of information from B to A – for example, any answer made by B to a question asked by A – unavoidably proceeds through physical means. Therefore it necessary takes the form of a measurement made by A on B.*"

That means that for every observer, anything outside of her own private perceptions (in particular any other observer) has to be treated as a physical system that can be entangled with other systems. It is then instructive to prove explicitly the fact that no conflict can arise between two observers on a simple example such as the measurement of the spin along Oz of an electron in an initial superposed state of spin. Suppose Bob has performed such a measurement on this electron. From Bob's point of view, a measurement has been made so he knows the value of the spin along Oz of this electron. According to the hanging-up mechanism Bob's consciousness is hung-up to one of the two possible branches "up" or "down". But from Alice's point of view, Bob is entangled with the electron, as described by the Schrödinger equation. Now Alice can perform the same measurement on the electron and Alice's consciousness will be hung-up as well to one of the two branches and will see one value. The crucial point is that this branch includes the state of Bob that is linked to the very same value. So when Alice, hung-up to that branch, speaks with Bob to know what Bob saw, she performs a measurement on Bob and, in accordance with the hanging-up mechanism, she cannot hear Bob saying anything else than the value that she has got herself. Alice will never hear Bob saying that he saw "up" when she saw "down". No conflict is possible and the intersubjectivity is preserved. Indeed this reasoning can be done for any other situation where Alice would like to compare her results with Bob's ones. Of course, we immediately want to know if it is possible that Bob saw "up" and Alice "down" even if Alice will never know. Actually, because of the relativity of states, this question is meaningless because any question must be stated from a specific observer point of view and there is no observer able to witness simultaneously what Alice and Bob saw independently. I refer the reader to [16] for a more detailed discussion of this point.

The image given by Convivial Solipsism is that our perceptions (our empirical reality) are built from "something" described through a universal entangled state vector which is relative to each observer whose consciousness is able to perceive only a part of it (the branch it is hung-up to). Because of the relativity of states and because each observer is hung-up to only one branch, there is no more any common shared reality, this is the reason why I call this interpretation "solipsism" but there is no way for two observers to notice any conflict between their perceptions, that is why it is a Convivial Solipsism (and also because unlike the usual solipsism, it allows for the existence of other conscious observers). A consequence is that some usual questions about the comparison between what two observers *really* saw become meaningless since each sentence must be understood as relative to a specific observer.

### 6. A new point of view for the delayed choice quantum eraser experiment

We can now come back to the EPR situation and the delayed choice quantum eraser experiment to understand what happens. Let's start by the EPR situation.

### 6.1. EPR situation

As we saw in §3, two particles A and B in a singlet state are measured by two spatially separated experimenters Alice and Bob. The state is given by equation (6):

$$|\psi\rangle = \tfrac{1}{\sqrt{2}}\left[\,|+\rangle^A|-\rangle^B - |-\rangle^A|+\rangle^B\,\right]$$

In the usual interpretation, when Alice does her measurement first, Bob finds the opposite outcome and it is usually stated that the collapse caused by Alice's measurement on A instantaneously results in a collapse of the state vector of B. Because of Bell's inequality, it is not possible to assume that the result is determined from the moment the particles separate. Moreover if the two measurements are space-like separated no one can be said to be before the other in an absolute way and so it becomes very difficult to assume that one of the two measurements causes the result of the other.

As we saw it is generally assumed that quantum mechanics is not local but that there is no way to use the collapse to communicate instantaneously. That seems to reassure many physicists but that is not very satisfactory. Indeed when summoned to analyze more precisely what that means physicists take refuge behind the fact that correlation is not causality. But as I mentioned above, this is not an acceptable reason since the difference between correlation and causality is the fact that a common cause can be invoked which is here forbidden by Bell's inequality. So, there is no real understanding of what happens during the measurement of distant entangled systems if we stay inside a realist framework assuming that state vectors represent the physical states of systems and that the collapse is a physical change of the state.

Let's analyze more precisely what happens in this situation. The reasoning goes like that: when Alice asks Bob which result he got and hears "down", she deduces in retrospect that the value has really been "down" as soon as Bob did his measurement. This is a reasonable deduction, but notice that she has no way to witness what happened at the exact moment when Bob did his measurement. So, is there a real obligation to assume that the value has really been "down" as soon as Bob did his measurement? Of course, in a realist framework where there is a unique reality which is the same for all the observers, this seems mandatory. Indeed the very fact that Bob did his measurement is supposed to imply that the value has been determined in an absolute way at that precise moment. Hence, assuming that the value has not been "down" at the moment Bob did his measurement would mean that the value has changed between the moment when Bob did his measurement and the moment when Alice asks Bob. Even worst, that would mean that Bob, when he answers Alice's question, gives an answer which is not conform to what he saw when he did his measurement. As we give credit here to Bob not to be a liar, that would mean that Bob's memory has spontaneously changed (Bob remembers having seen "down" whereas he actually saw "up"). All these things are highly improbable! So in a realist framework, this leads strongly to endorse Alice's deduction about the past and to support the fact that the value has really been "down" as soon as Bob did his measurement. Hence, that raises the problem of an instantaneous physical action at a distance as we said before. But things are different in Convivial Solipsism since the reality is relative to each observer and, more important, what an observer perceives of a system after a measurement is not something that has a physical impact on the system itself. It is then possible to understand what happens in this situation without any need of instantaneous physical action at a distance. Indeed, it is only when Alice asks Bob (in the future of her measurement on A) which value he found on B that she performs a measurement on Bob and learns what result he got on his particle. According to the hanging-up mechanism, she will necessarily hear "down" because she is hung-up to the branch which is linked to Bob having got this result in agreement with her own measurement on the first particle. Similarly if she performs a measurement on Bob's particle, she will find the same "down" value. But in Convivial Solipsism that does not mean that the second particle "was already in the state "down" before she asks Bob". Alice's deduction about the past, that seems inescapable in a realist framework, is no more valid. Indeed, the hanging-up mechanism is nothing else than the fact that Alice's consciousness hangs-up to the branch corresponding to the value "up" for her particle and "down" for Bob's particle, while the state vectors of the systems remain unchanged. The deduction about the "real value in the past" determined by Bob's measurement is meaningless. These posterior measurements done by Alice (on Bob and his particle) are of course time-like separated with Alice's first measurement and moreover, they do not affect physically the state of Bob's particle since, on the physical point of view, everything remains superposed.

It is now also easy to understand why the order of measurements does not matter. None is the physical cause of the other. There is no physical change. Since everything that can be said about correlations is relative to a specific observer, once she is hung-up to one branch by the first measurement, she will necessarily witness the correlations predicted by this branch for all subsequent measurements which will anyway be time-like separated. There is no way for an observer to make space-like separated measurements, any measurement is attached to a specific observer and a measurement has no physical effect on the system. There is no more mystery.

### 6.2. Delayed choice quantum eraser revisited

At the end of §4.3, I gave a description of the delayed choice quantum eraser experiment through equation (11):

$$|\psi\rangle = \frac{1}{2\sqrt{2}} |1\rangle^{\searrow} [|U\rangle + |L\rangle]^{\nearrow} + \frac{1}{2\sqrt{2}} |2\rangle^{\searrow} [|U\rangle + |L\rangle]^{\nearrow} + \frac{1}{2} |3\rangle^{\searrow} |L\rangle^{\nearrow} + \frac{1}{2} |4\rangle^{\searrow} |U\rangle^{\nearrow}$$

I emphasized the similitude with the EPR situation which is described by equation (6):

$$|\psi\rangle = \frac{1}{\sqrt{2}} \left[ |+\rangle^{A} |-\rangle^{B} - |-\rangle^{A} |+\rangle^{B} \right]$$

The explanation we gave above is that, when the wave function is written in this form, it is easy to see that if $D_1$ or $D_2$ clicks, the wave function of the signal photon collapses on $\frac{1}{\sqrt{2}} [|U\rangle + |L\rangle]^{\nearrow}$ and this will give rise to interference, while if $D_3$ (resp. $D_4$) clicks, the wave function of the signal photon collapses

on $|L\rangle^↗$ (resp. $|U\rangle^↗$) and there will be no interference. As I pointed out, this explanation uses the standard way to see the collapse and is subject to the same problems that those we raised about the EPR situation. But we can now analyze this experiment in the same way that the EPR situation, using the Convivial Solipsism framework. This is just a bit more complicated. Let's replace the down-arrow and the up-arrow by A and B respectively and assume that Alice is doing the measurement on the idler photon while Bob is doing the measurement on the signal photon.

$$|\psi\rangle = \frac{1}{2\sqrt{2}} |1\rangle^A [|U\rangle + |L\rangle]^B + \frac{1}{2\sqrt{2}} |2\rangle^A [|U\rangle + |L\rangle]^B + \frac{1}{2} |3\rangle^A |L\rangle^B + \frac{1}{2} |4\rangle^A |U\rangle^B \quad (12)$$

Actually, the fact that there are two observers does not matter because in Convivial Solipsism for any observer, another observer plays exactly the same role than an apparatus and must be treated similarly. So let's analyze the experiment from Alice's point of view. We can see the four detectors as a unique global apparatus intended to measure an observable with four possible results. When Alice does a measurement on an idler photon, she gets one of the four possible results. That means that her awareness hangs-up to the corresponding branch while the global state vector remains entangled. Now the experiment is using a series of successive measurements on different photons. Let's show explicitly what happens for two successive photons. The two photons are independent so the state vector including these two photons is the product of their state vectors which are each one of the form $|\psi\rangle$ given above.

$$|\psi\rangle_1 = \frac{1}{2\sqrt{2}} |1\rangle_1^A [|U\rangle + |L\rangle]_1^B + \frac{1}{2\sqrt{2}} |2\rangle_1^A [|U\rangle + |L\rangle]_1^B + \frac{1}{2} |3\rangle_1^A |L\rangle_1^B + \frac{1}{2} |4\rangle_1^A |U\rangle_1^B \quad (13)$$

$$|\psi\rangle_2 = \frac{1}{2\sqrt{2}} |1\rangle_2^A [|U\rangle + |L\rangle]_2^B + \frac{1}{2\sqrt{2}} |2\rangle_2^A [|U\rangle + |L\rangle]_2^B + \frac{1}{2} |3\rangle_2^A |L\rangle_2^B + \frac{1}{2} |4\rangle_2^A |U\rangle_2^B \quad (14)$$

$$|\psi\rangle = |\psi\rangle_1 |\psi\rangle_2$$

After the measurement on the two photons, Alice's awareness will be hung-up to the branch that is the product of the branches to which it has hung-up during each measurement. Assume for example that Alice gets the result 1 for the first measurement and 4 for the second one. Alice's awareness will be hung-up to: $\left\{\frac{1}{\sqrt{2}} |1\rangle_1^A [|U\rangle + |L\rangle]_1^B\right\} \left\{|4\rangle_2^A |U\rangle_2^B\right\}$. This process must be repeated for each couple of entangled photons included in the experiment.

When Alice learns about the results concerning the signal photons (directly through the records of these measurements or through questions she asks to Bob), this must not be interpreted as Alice learning a certain state of affairs already fixed but as Alice making a measurement on signal photons. Then, according to the hanging-up mechanism, she can get results coming only from the branch she is hung-up to. Hence for all the signal photons $j$ corresponding to idler photons for which she got the result 1, she is hung-up to $\left\{\frac{1}{\sqrt{2}} |1\rangle_j^A [|U\rangle + |L\rangle]_j^B\right\}$ and so she gets interference when she measures them (idem for result 2). For all the signal photons $k$ corresponding to idler photons for which she got the result 4, she is hung-up to $\left\{|4\rangle_k^A |U\rangle_k^B\right\}$ and so she gets no interference (idem for 3). That does not mean that there has been any physical effect in the past on the signal photons due to the measurement that she did on the idler photon. That only means that Alice's perception is confined into a branch that is determined by the first measurement she does on one part of an entangled composed system. There is no physical change because the system itself remains in an entangled state. That is also a way to understand better why the order of measurements does not matter. In a branch where a certain state of one part is linked to a certain state of the other part, it does not matter if you get first a result on one part or on the other. In both cases, the correlation will be the same.

Here also, as in the simpler EPR situation, using the Convivial Solipsism framework, we get an explanation that does not need any backward in time effect. Moreover, allowing to understand what the collapse is and when it occurs, this explanation removes all the paradoxes we faced in the usual interpretation.

## 7. Summary and conclusion

I have shown first that some often proposed presentations of the delayed choice experiments on one particle (i.e. without entangled particles) are misleading. In a standard framework where no issue is raised about the measurement problem and where the collapse of the state vector is accepted as such, these experiments can be explained in a straightforward way and do not need to appeal to backward in time effects. A proper analysis shows that the double slit experiment that Feynman considered as containing the main mystery of quantum mechanics can be understood easily and that the possibility of a delayed choice brings no additional reasons to question the standard formalism about the concept of measurement. These delayed choice experiments can also be explained readily and add nothing to the questions that can already be raised from the mere conflict between the Schrödinger equation and the reduction postulate.

The situation is different when switching to more complicated experiments involving entangled particles. My goal in this paper has been to show that the paradoxes raised by considering that the collapse is a real physical action changing instantaneously the state of the system, already present in the EPR situation, are emphasized in the delayed choice quantum eraser (while being essentially of the same nature). The simple explanation provided in the case of delayed choice experiments involving only one particle does not work anymore. The main reason is that, when there is only one particle, it is possible to understand what happens either without any modification of the state vector of the particle (in the delayed choice double slit experiment) or with a modification that can be attributed to a clear physical effect (the interaction with the beam splitter when it is present in the delayed choice Mach-Zender experiment). Whereas in the EPR situation and in the delayed choice quantum eraser experiment, the correlations found after the two measurements seem to imply that the state vector of the distant entangled system depends on the measurement of the first system in a way that cannot be explained by a normal physical effect. So, if this state vector is considered as representing the "real" state of the system and if the measurement is assumed to change it (this is the case in a realist framework), one is in trouble for explaining how and why it has changed. On the contrary, if this state vector is considered as something that is relative to the observer while the measurement has no physical effect on it and everything becomes clear. This latter position is the solution proposed by Convivial Solipsism.

Quantum mechanics forbids to keep unchanged the classical picture of an external and independent world where things happen by themselves and where an observation is merely the fact for an observer to notice something that has already happened. In his seminal paper "Law without law" [24] it is interesting to see that Wheeler, while abandoning partly this way of thinking, still sticks to a point of view leading him to ask many puzzling questions about "reality". Taking the example of the light coming from a very distant quasar, he discusses the status of reality of these photons. On the one hand, he denies that we have the right to speak of what the photons did before they were observed ("*no elementary phenomenon is a phenomenon until it is an observed phenomenon*") and he argues for the fact that "*we, here, now, have a part in bringing about that which had already happened at a time when no observer existed*". On the other hand, he is worried about denying any reality to the "*more numerous relict photons that escape our telescope*" and is led to conclude that "*their reality is of a paler and more theoretic hue*". It is easy to see that, reluctant to abandon totally the concept of an external independent reality resembling more or less to the reality we are accustomed to, but nevertheless willing to account for the quantum facts, he is prisoner of a network of incompatible intuitions.

Now, it is interesting to compare the image given by Convivial Solipsism with the one Wheeler had in mind [24]:

"*What keeps these images of something "out there" from degenerating into separate and private universes: one observer, one universe; another observer, another universe? That is prevented by the very solidity of these iron posts, the elementary acts of observership-participancy. That is the importance of Bohr's point that no observation is an observation unless we can communicate the results of that observation to others in plain language.*"

In Convivial Solipsism we are led to a more drastic conclusion:

Nothing keeps these images of something "out there" from degenerating into separate and private universes: one observer, one universe; another observer, another universe. That is actually exactly what happens, as strange as it may seem. The importance of Bohr's point that no observation is an observation unless we can communicate the results of that observation to others in plain language only prevents that observers become aware of the fact that their universes are perhaps different. This is so because any communication between two observers is nothing else than a measurement of one observer by the other and that the rules of quantum mechanics prevent any discrepancy to be noticed.